\documentclass[%
 reprint,
superscriptaddress,
 amsmath,amssymb,
 aps,
]{revtex4-2}
\usepackage{graphicx} 
\usepackage{tikz}
\usepackage{natbib}%
\usepackage{hyperref}
\usepackage{subfigure}
\usepackage{amsmath}
\usepackage{comment}

\usepackage{svg}
\usepackage{booktabs}
\usepackage{multirow}
\usepackage{array}
\usepackage{makecell}

\begin{document}

\title{Strongly Coupled Exciton - Hyperbolic-phonon-polariton Hybridized States in hBN-encapsulated Biased Bilayer Graphene}

\author{Tomer Eini}
\affiliation{School of Electrical Engineering, Faculty of Engineering, Tel Aviv University, Tel Aviv 6997801, Israel}
\author{N. M. R. Peres}
\affiliation{International Iberian Nanotechnology Laboratory (INL), Av. Mestre José Veiga, 4715-330, Braga, Portugal}
\affiliation{Centro de Física das Universidades do Minho e do Porto (CF-UM-UP) e Departamento de Física, Universidade do Minho, P-4710-057 Braga, Portugal}
\affiliation{POLIMA—Center for Polariton-driven Light-Matter Interactions, University of Southern Denmark, Campusvej 55, DK-5230 Odense M, Denmark}
\author{Yarden Mazor}
\affiliation{School of Electrical Engineering, Faculty of Engineering, Tel Aviv University, Tel Aviv 6997801, Israel}
\author{Itai Epstein}
\email{itaieps@tauex.tau.ac.il}
\affiliation{School of Electrical Engineering, Faculty of Engineering, Tel Aviv University, Tel Aviv 6997801, Israel}
\affiliation{Center for Light-Matter Interaction, Tel Aviv University, Tel Aviv 6997801, Israel}
\affiliation{QuanTAU, Quantum Science and Technology Center, Tel Aviv University, Tel Aviv 6997801, Israel}


\begin{abstract}
Excitons in biased bilayer graphene are electrically tunable optical excitations residing in the mid-infrared (MIR) spectral range, where intrinsic optical transitions are typically scarce. Such a tunable material system with an excitonic response offer a rare platform for exploring light–matter interactions and optical hybridization of quasiparticles residing in the long wavelength spectrum. In this work, we demonstrate that when the bilayer is encapsulated in hexagonal-boron-nitride (hBN)—a material supporting optical phonons and hyperbolic-phonon-polaritons (HPhPs) in the MIR—the excitons can be tuned into resonance with the HPhP modes. We find that the overlap in energy and momentum of the two MIR quasi-particles facilitate the formation of multiple strongly coupled hybridized exciton-HPhP states. Using an electromagnetic transmission line model, we derive the dispersion relations of the hybridized states and show that they are highly affected and can be manipulated by the symmetry of the system, determining the hybridization selection rules. Our results establish a general tunable MIR platform for engineering strongly coupled quasiparticle states in biased graphene systems, opening new directions for studying and controlling light–matter interactions in the long-wavelength regime.
\end{abstract}

\maketitle

\section{Introduction}

Polaritons are quasiparticles formed by the hybridization of photons with a material excitation, and thus have attributes associated with both light and matter. In semiconductors, such hybridization can occur when photons interact with a bound electron-hole pair in the form of excitons, giving rise to exciton-polaritons. The phenomena can be described in the formalism of strong coupling between the two quasiparticles, resulting in the superposition of the states into two new eigenmodes, characterized by a Rabbi splitting and its resulting anti-crossing \cite{Weisbuch1992ObservationMicrocavity, Kavokin2010Exciton-polaritonsPerspectives, Deng2010Exciton-polaritonCondensation, Timofeev2012ExcitonFrontiers, Yamamoto1999MesoscopicOptics, Liu2014StrongCrystals, Schneider2018Two-dimensionalCoupling}. Exciton-polaritons formed by an exciton interacting with a cavity photon have been extensively researched, providing an exemplary system for such strong coupling, and have enabled the demonstration of phenomena such as Bose–Einstein condensates, polariton amplification, superfluidity, and more \cite{Byrnes2014ExcitonpolaritonCondensates, Deng2010Exciton-polaritonCondensation}.

Generally, any two states overlapping in energy and momentum can interact in a similar manner, whether these are plasmonic modes, emitters, molecules in cavities, and others \cite{Deng2010Exciton-polaritonCondensation, Tormo2014StrongReview, Nagarajan2021ChemistryCoupling, Chikkaraddy2016Single-moleculeNanocavities, Dovzhenko2018LightmatterApplications, Liu2014StrongCrystals, Luo2024StrongMaterials}.
While the interacting states can be quantum, classical analogs can also exhibit such strong and weak coupling regimes \cite{Novotny2010StrongPerspective}. 

Recently, Bernal-stacked biased bilayer graphene (BBLG) has been demonstrated to support electrically tunable excitons in the mid-infrared (MIR) spectrum \cite{Park2010TunableGraphene, Ju2017TunableGraphene, Ju2020UnconventionalGraphene, Duarte2024MoirePressure, Henriques2022AbsorptionGraphene, Quintela2022TunableGraphene, Eini2025ElectricallyGrapheneb}. While unbiased bilayer graphene is semi-metallic, inversion symmetry can be broken by applying an electric field across the graphene layers, for example, resulting in the opening of a bandgap \cite{Castro2007BiasedEffect, Zhang2009DirectGraphene., Oostinga2007Gate-inducedDevices}, and the ability to support excitons that can be spectrally tuned with the applied bias \cite{Park2010TunableGraphene, Ju2017TunableGraphene, Ju2020UnconventionalGraphene, Duarte2024MoirePressure, Henriques2022AbsorptionGraphene, Quintela2022TunableGraphene, Eini2025ElectricallyGrapheneb}. These excitons can also interact with photons, giving rise to intriguing polaritonic phenomena, such as tunable graphene-exciton-polaritons \cite{Eini2025ElectricallyGrapheneb}, or cavity exciton-polaritons when the BBLG is placed within an optical cavity \cite{Berman2012SuperfluidityMicrocavity, DeLiberato2015PerspectivesPolaritonics,Duarte2025TunableGraphene}.

It is important to note that excitons in BBLG have only been observed so far in BLG samples that were encapsulated with hexagonal-boron-nitride (hBN), which plays an important role in obtaining high-quality electrical and optical response in 2D materials \cite{Wang2013One-dimensionalMaterial, Epstein2020Far-fieldVolumesb, Epstein2020Near-unityCavityc, Woessner2014HighlyHeterostructures, Ni2018FundamentalPlasmonics, Lundeberg2017TuningPlasmonics, Iranzo2018ProbingHeterostructure, Hesp2021ObservationGraphene}. On top of these attributes, hBN possesses interesting optical properties by itself, such as a hyperbolic material response and highly-confined hyperbolic-phonon-polaritons (HPhPs) with low propagation losses \cite{Dai2014TunableNitride, Fali2019RefractivePropagation, Dai2019HyperbolicNitride, Hu2020PhononMaterials, Ni2021Long-LivedMaterials, Li2015HyperbolicFocusing, Jacob2014HyperbolicPhononpolaritons, Giles2016ImagingNitride, Klein2025Nanometer-ScaleResonatorsb}.
These HPhPs are supported in two spectral ranges at MIR frequencies, between the two transverse optical (TO) and the longitudinal optical (LO) phonons of hBN, known as "Reststrahlen bands". The hBN-encapsulated BBLG system can thus support various excitonic and phononic quasi-particles, which can interact strongly with light, providing a rich playground for studying the formation of new hybridized states in the long wavelength spectral range.

In this work, we unveil the existence of electrically tunable hybridized MIR exciton-phonon polariton modes in the BBLG system. We show that BBLG excitons can be tuned into resonance with hBN HPhP modes, and find multiple strongly coupled hybridized exciton-HPhP states. We derive the dispersion relations of these hybridized states using an electromagnetic transmission line model and show that they are highly affected and can be manipulated by the symmetry and parameters of the system, determining the hybridization selection rules. These results establish a general tunable MIR platform for engineering strongly coupled quasiparticle states in additional biased graphene systems, such as rhombohedral trilayer graphene for example, opening new directions for studying and controlling light–matter interactions in the long-wavelength regime.


\section{Optical properties of HPhPs and BBLG excitons}
\label{section cond}
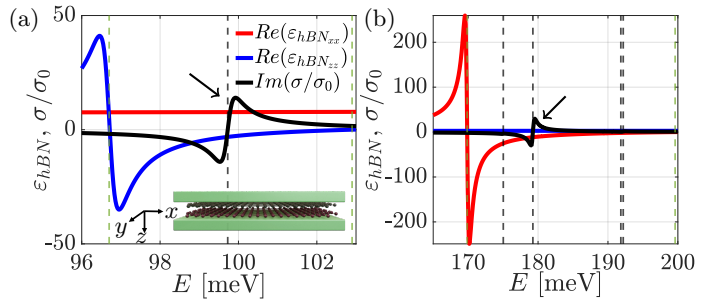
\begin{figure}[t]
    \centering
    \hspace*{-13pt} 
    \begin{tikzpicture}
    \node[anchor=south west,inner sep=0] (image) 
    at (0,0) 
    {\includegraphics[width=1.03\columnwidth]{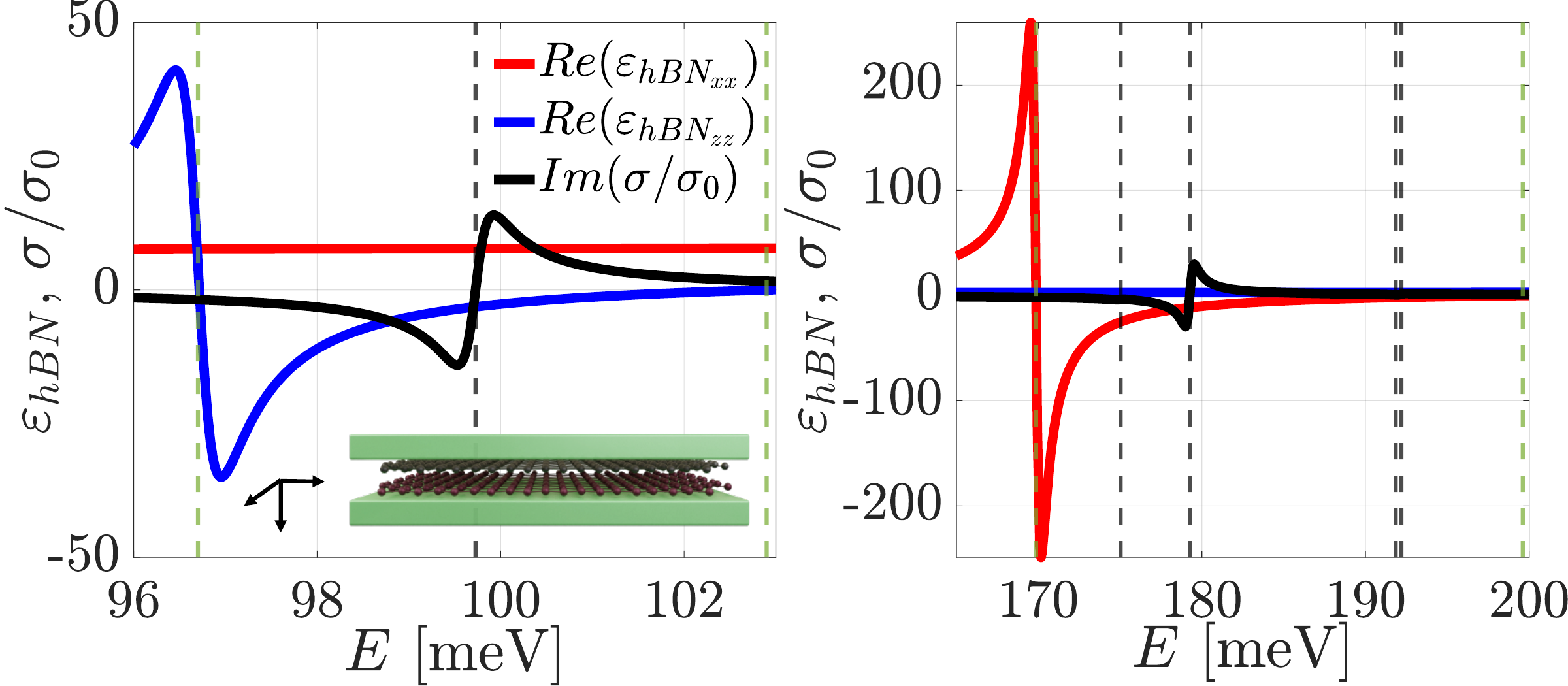}};
    \begin{scope}[x={(image.south east)},y={(image.north west)}]
    \node at (0,0.95) {(a)};
    \node at (0.53,0.95) {(b)};
    \node at (0.175,0.21) {$z$};
    \node at (0.22,0.295) {$x$};
    \node at (0.145,0.235) {$y$};
    \end{scope}
    \draw[->, thick, black] (2.2, 3) -- (2.6, 2.7);
    \draw[->, thick, black] (7.2, 2.7) -- (6.9, 2.4);

    \end{tikzpicture}
    \caption{Conductivity of BBLG for (a) $V=58\,\mathrm{meV}$ and (b) $V=115\,\mathrm{meV}$, with the in-plane and out-of-plane permittivity of hBN. The main (marked with arrows) and additional exciton energies (dashed black lines) are inside the band limits (dashed green lines). The configuration of BBLG encapsulated by hBN is illustrated as an inset.}
    \label{fig_cond}
\end{figure}

The systems under investigation is the hBN-encapsulated BBLG (Fig. \ref{fig_cond}), where its bandgap has been opened via an applied bias via two electrodes placed above and below the encapsulating hBN, with the bandgap size being controlled by the magnitude  of the applied bias \cite{Henriques2022AbsorptionGraphene, Quintela2022TunableGraphene, Park2010TunableGraphene, Ju2017TunableGraphene, Ju2020UnconventionalGraphene, Eini2025ElectricallyGrapheneb}. Taking into account the presence of excitons in the system the optical conductivity of the BBLG becomes \cite{Henriques2022AbsorptionGraphene,Quintela2022TunableGraphene, Eini2025ElectricallyGrapheneb, Pedersen2015IntrabandGeneration,Aversa1995NonlinearAnalysis,E-PeriodicaElectrodynamics, Lehmann1954UberFelder, Quintela2024TunableGraphene}:
\begin{equation} \label{Eq conductivity}
   \sigma=4i\sigma_0\sum_{n} \frac{f_n}{E -E_n+i\frac{\Gamma_n}{2}},
\end{equation}
where $\sigma_0=\frac{e^2}{4\hbar}$ is the universal conductivity of graphene, $E$ is the energy, $f_n$, $E_n$ and $\Gamma_n$ are the oscillator strength, exciton energy, and non-radiative decay rate of the excitonic resonance of the $n$th exciton Rydberg series. 

The optical response of hBN is commonly described by its frequency-dependent permittivity \cite{Dai2014TunableNitride}: 
\begin{equation}
    \varepsilon_{hBN_{j}} = \varepsilon_{hBN_{j}^\infty} \bigg (1 + \frac{\omega_{LO,j} ^2 -\omega_{TO,j} ^2}{\omega_{TO,j} ^2 - \omega ^2 -i \omega \Gamma_j} \bigg),
    \label{Eq hBN perm}
\end{equation}

where $j=xx,zz$ is the in-plane or out-of-plane direction, $\varepsilon_{hBN_{j}^\infty}$ is the background permittivity, $\omega_{LO,j}$ and $\omega_{LO,j}$ are the longitudinal and transverse optical phonon frequencies, respectively, and $\Gamma_j$ is the phonon decay rate. Each permittivity component, separately, is negative in a different spectral region of the optical response, referred to as the Reststrahlen band \cite{Caldwell2014Sub-diffractionalNitride}. Therefore, each Reststrahlen band satisfies the condition $Re(\varepsilon_{hBN_{xx}}) \cdot Re(\varepsilon_{hBN_{zz}})<0$, making the hBN a hyperbolic material in these spectral ranges \cite{Poddubny2013HyperbolicMetamaterials, Shekhar2014HyperbolicApplications, Ferrari2015HyperbolicApplications}. This condition, leads to a dispersion relation in the form of a hyperboloid isofrequency surface in $k$ space, enabling the material to support HPhPs, which have the following dispersion relation for an air/hBN slab/air structure (appendix \ref{section app HPhP}, \cite{Dai2014TunableNitride, Fali2019RefractivePropagation, Dai2019HyperbolicNitride, Hu2020PhononMaterials, Ni2021Long-LivedMaterials, Li2015HyperbolicFocusing, Jacob2014HyperbolicPhononpolaritons, Giles2016ImagingNitride}):

\begin{equation}    
    q_{HPhP}(d)=\frac{i}{d} \sqrt{\frac{\varepsilon_{hBN_{zz}}}{\varepsilon_{hBN_{xx}}}} \bigg(2 \arctan \bigg(\frac{i}{\varepsilon_{hBN_{eff}}} \bigg)+\pi L \bigg),
    \label{Eq HPhP}
\end{equation}
where $q_{HPhP}$ is the component of the wavevector in the
$x$ direction (the momentum), $d$ is the hBN thickness, $\varepsilon_{hBN_{eff}}=\sqrt{\varepsilon_{hBN_{xx}}\varepsilon_{hBN_{zz}}}$ is the effective permittivity of hBN \cite{Woessner2014HighlyHeterostructures} and $L$ is an integer signifying the modal order:  $L=-1,-2,-3,...$ for the lower Reststrahlen band and $L=0,1,2,...$ for the upper Reststrahlen band.

Figure \ref{fig_cond} presents the permittivity components of hBN in the lower (a) and upper (b) Reststrahlen bands together with the conductivity of BBLG for bias of $V=58meV$ (a) and $V=115meV$ (b). In the lower Reststrahlen band, the out-of-plane permittivity of hBN is negative while the in-plane is positive, and vice versa for the upper Reststrahlen band. The conductivity of the BBLG has excitonic resonances within these spectral bands, where the main resonance (with the largest oscillator strength) change signs of imaginary part around its exciton energy.

\section{Exciton-HPhP Hybridization}\label{section hybridization}

The bias dependency of the conductivity and excitonic resonances has previously been studied 
\cite{Henriques2022AbsorptionGraphene,Quintela2022TunableGraphene, Eini2025ElectricallyGrapheneb}, enabling to shift the BBLG excitons' resonance into the hBN's Reststrahlen bands (Fig. \ref{fig_cond}), allowing for the possible hybridization of the BBLG  excitons with hBN's HPhP. For studying the possible hybridized response in a convenient manner, we assume the system under investigation to be air/hBN/BBLG/hBN/air structure while setting the bias via the conductivity in Eq. \ref{Eq conductivity} \cite{Eini2025ElectricallyGrapheneb} rather than simulating it through the presence of gate electrodes. Then, using a transmission line model (appendix \ref{Section app TL}, \cite{DavidkChengField-and-wave-electromagnetics}), we can find the dispersion relation of the system (appendix \ref{Section app hyb}).

\subsection{Symmetric structure}
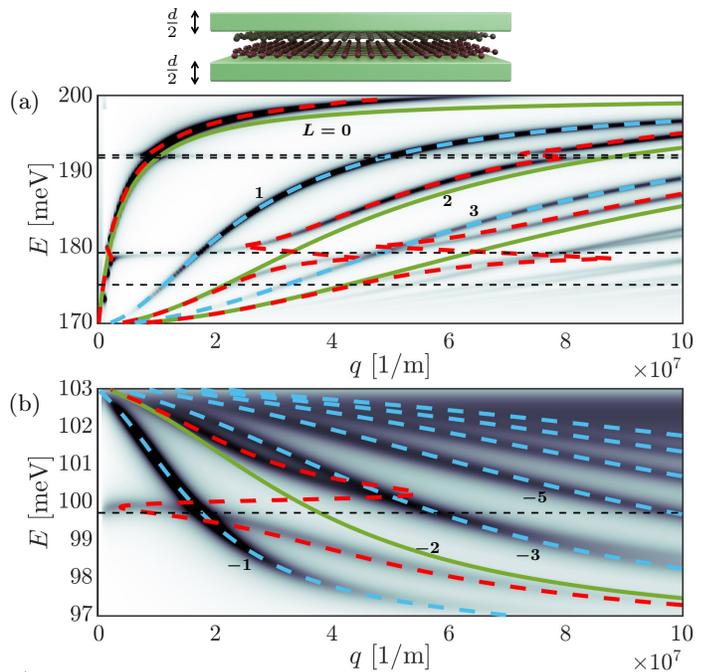
\begin{figure}[t]
    \centering
    \hspace*{-13pt} 
    \begin{tikzpicture}
    \node[anchor=south west,inner sep=0] (image) 
    at (0,0) 
    {\includegraphics[width=1.03\columnwidth]{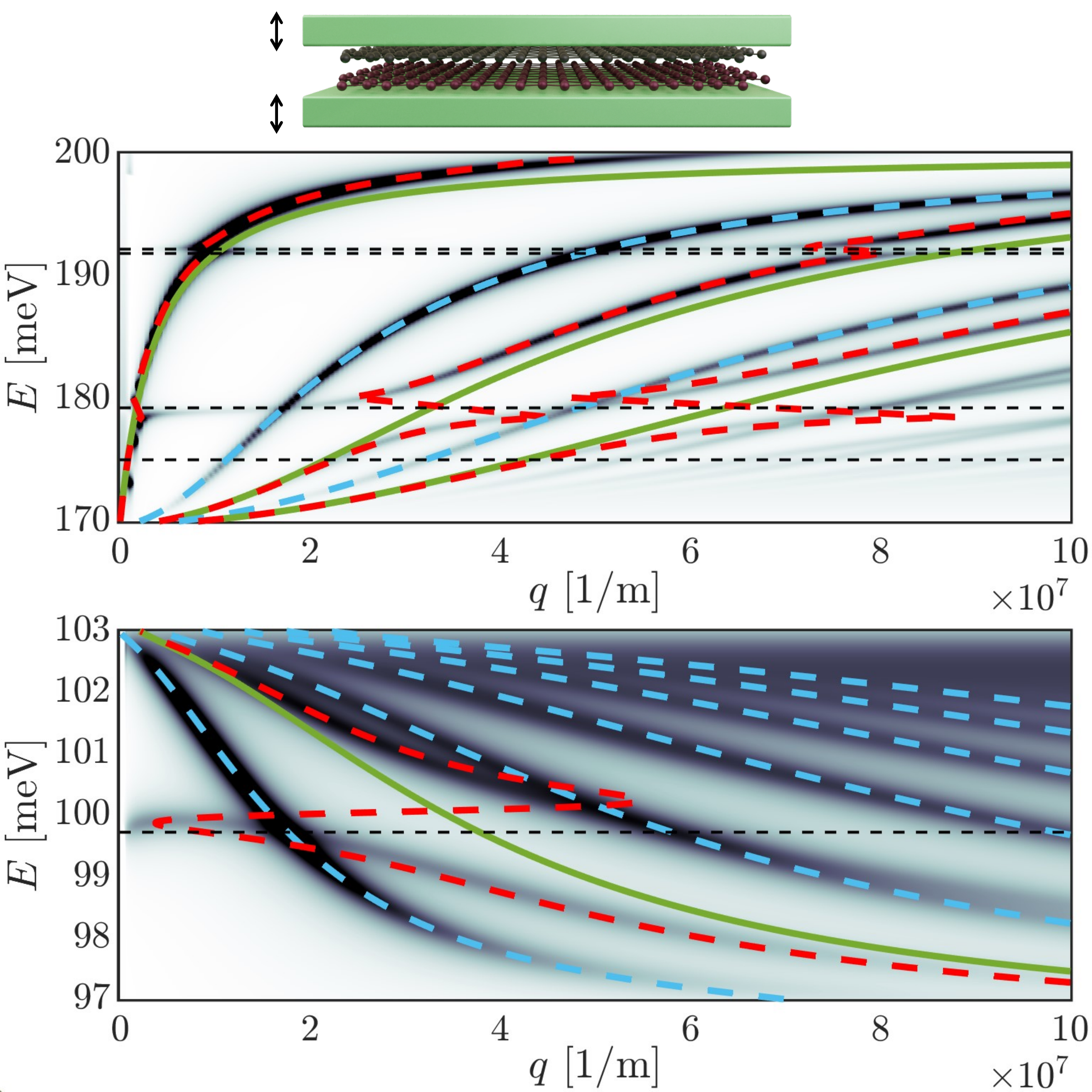}};
    \begin{scope}[x={(image.south east)},y={(image.north west)}]
    \node at (0,0.85) {(a)};
    \node at (0,0.4) {(b)};
    \node at (0.22,0.97) {$\frac{d}{2}$};
    \node at (0.22,0.9) {$\frac{d}{2}$};
    \node [scale=0.7]  at (0.45,0.81) {$\boldsymbol{L=0}$};
    \node [scale=0.7]  at (0.35,0.716) {$\boldsymbol{1}$};
    \node [scale=0.7]  at (0.63,0.705) {$\boldsymbol{2}$};
    \node [scale=0.7]  at (0.67,0.69) {$\boldsymbol{3}$};
    \node [scale=0.7]  at (0.32,0.16) {$\boldsymbol{-1}$};
    \node [scale=0.7]  at (0.6,0.185) {$\boldsymbol{-2}$};
    \node [scale=0.7]  at (0.75,0.175) {$\boldsymbol{-3}$};
    \node [scale=0.7]  at (0.76,0.26) {$\boldsymbol{-5}$};
    \end{scope}
    \end{tikzpicture}
    \caption{Hybridization of BBLG excitons and hBN HPhPs for the symmetric structure, with $\frac{d}{2}=50\,\mathrm{nm}$. 
    The dispersion relation of the hybridized polariton in the symmetric case for BBLG with $V=115\,\mathrm{meV}$ (a) and $V=58\,\mathrm{meV}$ (b), calculated from Eq. \ref{Eq hybrid sym even} (dashed red lines) and Eq. \ref{Eq hybrid sym odd} (dashed blue lines), and simulated using TMM (colormap). Even modes of HPhP (green lines), calculated from Eq. \ref{Eq HPhP}, and exciton energies of BBLG (dashed black lines) are also plotted. The modal orders $L$ are marked in the figure. 
    The even hybridized modes present an anticrossing behavior between even modes of HPhPs and the excitonic resonances of BBLG while the odd hybridized modes have the dispersion relation of odd HPhPs.
    The configuration is illustrated above the figure.}
    \label{fig_hybrid_sym}
\end{figure}

To understand the physical behavior of the system, first we look at a symmetric structure with both hBN layers having a $\frac{d}{2}$ thickness (Fig. \ref{fig_hybrid_sym}). Owing to the symmetry around $z=0$ the solution can be divided into even and odd solutions in terms of the transverse electric field, similarly to HPhP in the air/hBN/air structure introduced in Eq. \ref{Eq HPhP} (appendix \ref{section app HPhP}).
The dispersion relation of the symmetric structure with $\frac{d}{2}=50\,\mathrm{nm}$ is presented in Fig. \ref{fig_hybrid_sym} (a) and (b) for the two hBN Reststrahlen bands.

The even solution is given by (appendix \ref{Section app sym hyb}):

\begin{align}    
    \begin{split}
    &
    q_{Hybrid}^{even}
    =\\
    & =
    \frac{i}{d+d_\sigma} \sqrt{\frac{\varepsilon_{hBN_{zz}}}{\varepsilon_{hBN_{xx}}}} \bigg(2 \arctan \bigg(\frac{i}{\varepsilon_{hBN_{eff}}} \bigg)+2 \pi L_{even} \bigg),
    \label{Eq hybrid sym even}
    \end{split}
\end{align}

where $d_\sigma=\frac{i \sigma}{\omega \varepsilon_0 \varepsilon_{hBN_{xx}}}$, which provides the additional polarization current that flows on the BBLG in terms of length (appendix \ref{Section app d_sigma}). It can be seen that the obtained dispersion relation in Eq. \ref{Eq hybrid sym even} is the dispersion relation of even modes of HPhP (Eq. \ref{Eq HPhP}), with an effective thickness of $d_{eff}=d+d_{\sigma}$, i.e.: $q_{Hybrid}^{even}=q_{HPhP}^{even} (d+d_{\sigma} )$.
The analytical solution for the even modes given by Eq. \ref{Eq hybrid sym even} is presented in Fig. \ref{fig_hybrid_sym} (a) and (b) with dashed red lines.

The odd solution is given by (appendix \ref{Section app sym hyb}):

\begin{align}    
    \begin{split}
    &
    q_{Hybrid}^{odd}
    =\\
    & =
    \frac{i}{d} \sqrt{\frac{\varepsilon_{hBN_{zz}}}{\varepsilon_{hBN_{xx}}}} \bigg(2 \arctan \bigg(\frac{i}{\varepsilon_{hBN_{eff}}} \bigg)+ \pi (2L_{odd}-1) \bigg).
    \label{Eq hybrid sym odd}
    \end{split}
\end{align}

The odd modes correspond to odd modes of HPhP with thickness of $d$: $q_{Hybrid}^{odd}=q_{HPhP}^{odd} (d)$.
The analytical solution for the odd modes given by Eq. \ref{Eq hybrid sym odd} is presented in Fig. \ref{fig_hybrid_sym} (a) and (b) with dashed blue lines.
Excellent agreement between the TMM simulation (colormap) and the analytical solution for the even (dashed red line) and odd modes (dashed blue lines), obtained from Eq. \ref{Eq hybrid sym even} and \ref{Eq hybrid sym odd}, can be seen in Fig. \ref{fig_hybrid_sym}.

From Fig. \ref{fig_hybrid_sym} it is clear that the even modes present an anti-crossing between the HPhP (green lines) and the exciton energies (dashed black lines). This anti-crossing, being a signature of the interaction as described in the introduction, is an evidence of the hybridization between the HPhPs and the excitons in the BBLG. Thus, due to the tunability of the excitons in the BBLG, hybridized exciton polaritons appear in both Reststrahlen bands.

The anticrossing between the even HPhPs and the main excitonic resonances can be understood intuitively by examining Eq. \ref{Eq hybrid sym even} (appendix \ref{Section app sym hyb}): the hybridized dispersion relation converges to the dispersion relation of HPhPs far from the excitonic resonances and experiences a decrease or increase momentum compared to HPhPs near the excitonic resonances. Unlike the main excitonic resonance, the additional resonances do not present exact anticrossings of the HPhPs dispersion around these resonances, although the general behavior is similar.
For instance, in Fig. \ref{fig_hybrid_sym} (a), the even hybridized mode (red curve) that corresponds to the even HPhP (green curve) with $L=2$ has lower momentum below and above $\approx 192\,\mathrm{meV}$.

 Note that in our case, the lower Reststrahlen band only supports one even mode, with modal order $L_{even}=-1$ which correspond to $L=2L_{even}=-2$ of HPhP (see Fig. \ref{fig_hybrid_sym} (b)), as explained in appendix \ref{Section app sym hyb}.

The odd modes, however, do not present an anti-crossing behavior as can be seen in Fig. \ref{fig_hybrid_sym}, since their transverse electric field is $0$ on the BBLG at $z=0$, yielding no surface current on the BBLG, $J_s=\sigma \mathbb{E}_x (z=0)=0$.
The transverse magnetic field at $z=0$ satisfies $\mathbb{H}_{y}(0^-)-\mathbb{H}_{y}(0^+)=J_s=0$, which is the same condition in air/hBN/air structure introduced in Eq. \ref{Eq HPhP}. Hence, the magnetic field is continuous and the odd modes are not affected by the BBLG (Eq. \ref{Eq hybrid sym odd}).

\subsection{Asymmetric structure}
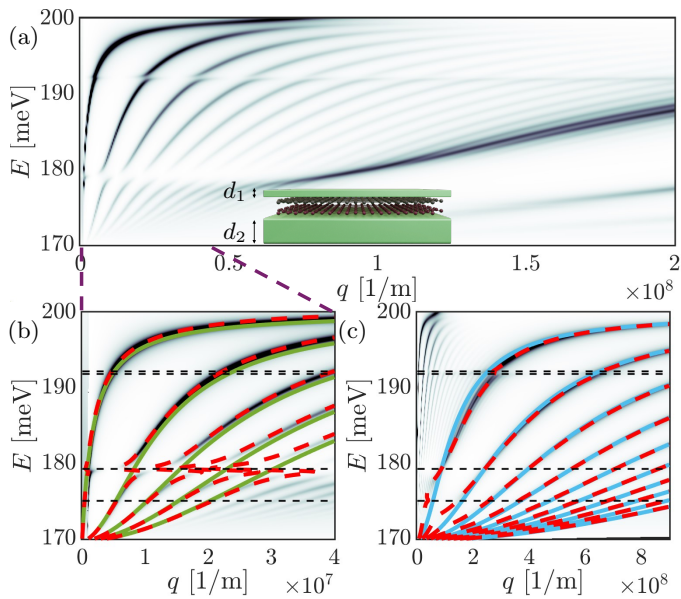
\begin{figure} [b]
    \centering
    \hspace*{-13pt} 
    \begin{tikzpicture}
    \node[anchor=south west,inner sep=0] (image) 
    at (0,0) 
    {\includegraphics[width=1.03\columnwidth]{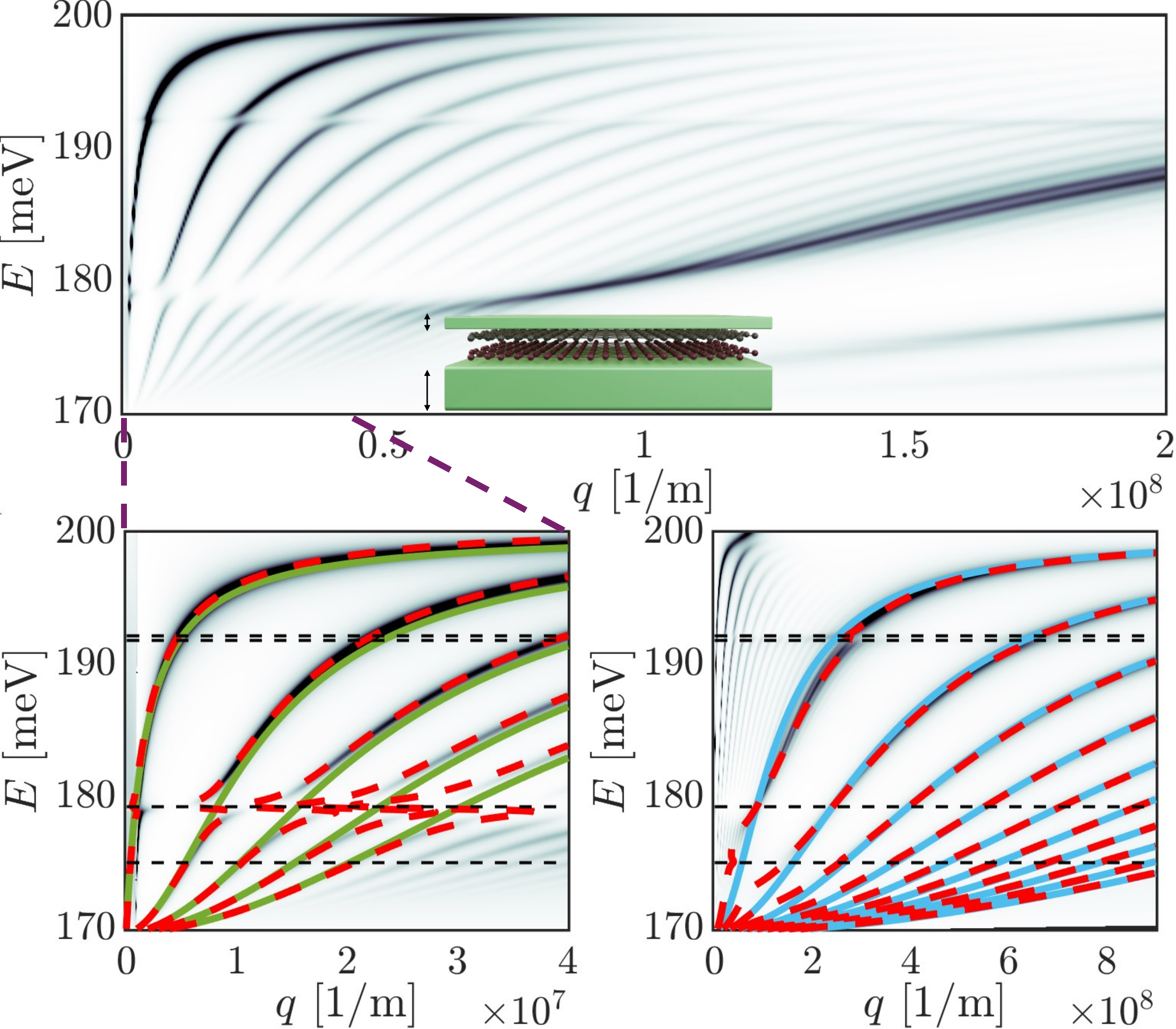}};
    \begin{scope}[x={(image.south east)},y={(image.north west)}]
    \node at (0.02,0.95) {(a)};
    \node at (0.02,0.45) {(b)};
    \node at (0.51,0.45) {(c)};
    \node at (0.335,0.69) {$d_1$};
    \node at (0.335,0.62) {$d_2$};
    \end{scope}
    \end{tikzpicture}
     \caption{Hybridization of BBLG excitons and hBN HPhP for the asymmetric structure in the upper Reststrahlen band, with $d_1=10\,\mathrm{nm}$ and $d_2=200\,\mathrm{nm}$. (a) The dispersion relation  of the hybridized polariton for BBLG with $V=115\,\mathrm{meV}$ simulated using TMM (colormap) showing two sets of modes. 
    (b) Zoom in on the low momentum modes of panel (a) together with the dispersion relation calculated from Eq. \ref{Eq hybrid asym low} (dashed red lines). Modes of HPhP for hBN with thickness of $d_1+d_2$ (green lines) and exciton energies of BBLG (dashed black lines) are also plotted.
    (c) Zoom out to the high momentum modes of panel (a) together with the dispersion relation calculated from Eq. \ref{Eq hybrid asym high} (dashed red lines). Odd modes of HPhP for hBN with thickness of $2d_1$ (blue lines) and exciton energies of BBLG (dashed black lines) are also plotted.
    The low momentum hybridized modes present an anticrossing behavior while the high momentum hybridized modes cross the main excitonic resonances. The configuration is illustrated as inset.}
    \label{fig_hybrid_asym_upper}
\end{figure}

Next, we look at the asymmetric structure with an arbitrary top and bottom hBN thicknesses $d_1$ are $d_2$ (Fig. \ref{fig_hybrid_asym_upper}). To gain a full understanding of the physical phenomena in the asymmetric structure we choose to analyze a case with an approximate analytical solution, where $d_1,|d_\sigma| \ll d_2$. Using a transmission line model, we can find two sets of solutions for the hybridized modes (appendix \ref{Section app asym hyb}). The dispersion relations of the asymmetric structure in the upper Reststrahlen band are presented in Fig. \ref{fig_hybrid_asym_upper}, with $d_1=10\,\mathrm{nm}$ and $d_2=200\,\mathrm{nm}$. Fig. \ref{fig_hybrid_asym_upper} (a) shows the existence of two sets of modes, distinguished by their lower and higher momentum.

The first set of modes, is the set of low momentum modes, which satisfy $|k_{hBN_z} d_{1,\sigma}| \ll 1$, where $k_{hBN_z}$ is the $z$ component of the wavevector in the hBN layers. The dispersion relation of these modes is given by (appendix \ref{Section app asym hyb}):

\begin{align}    
    \begin{split}
    &
    q_{Hybrid}^{low}
    =\\
    & =
    \frac{i}{d_1+d_2+d_\sigma} \sqrt{\frac{\varepsilon_{hBN_{zz}}}{\varepsilon_{hBN_{xx}}}} \bigg(2 \arctan \bigg(\frac{i}{\varepsilon_{hBN_{eff}}} \bigg)+ \pi L_{low} \bigg).
    \label{Eq hybrid asym low}
    \end{split}
\end{align}

The dispersion relation obtained in Eq. \ref{Eq hybrid asym low} is equivalent to the dispersion relation of HPhP with an effective thickness of $d_{eff}=d_1+d_2+d_\sigma$:  $q_{Hybrid}^{low}=q_{HPhP} (d_1+d_2+d_\sigma)$.
The analytical solution for the low momentum modes given by Eq. \ref{Eq hybrid asym low} is presented in Fig. \ref{fig_hybrid_asym_upper}  (b) with dashed red lines.
The low momentum hybridized modes in the asymmetric structure present the anticrossing that was obtained for the even hybridized modes in the symmetric structure. Here, all the modes experience anticrossing (Fig. \ref{fig_hybrid_asym_upper} (b)), since there is no nullification of the transverse electric field on the BBLG due to symmetry.

\begin{figure} [b]
    \centering
    \hspace*{-13pt} 
    \begin{tikzpicture}
    \node[anchor=south west,inner sep=0] (image) 
    at (0,0) 
    {\includegraphics[width=1.03\columnwidth]{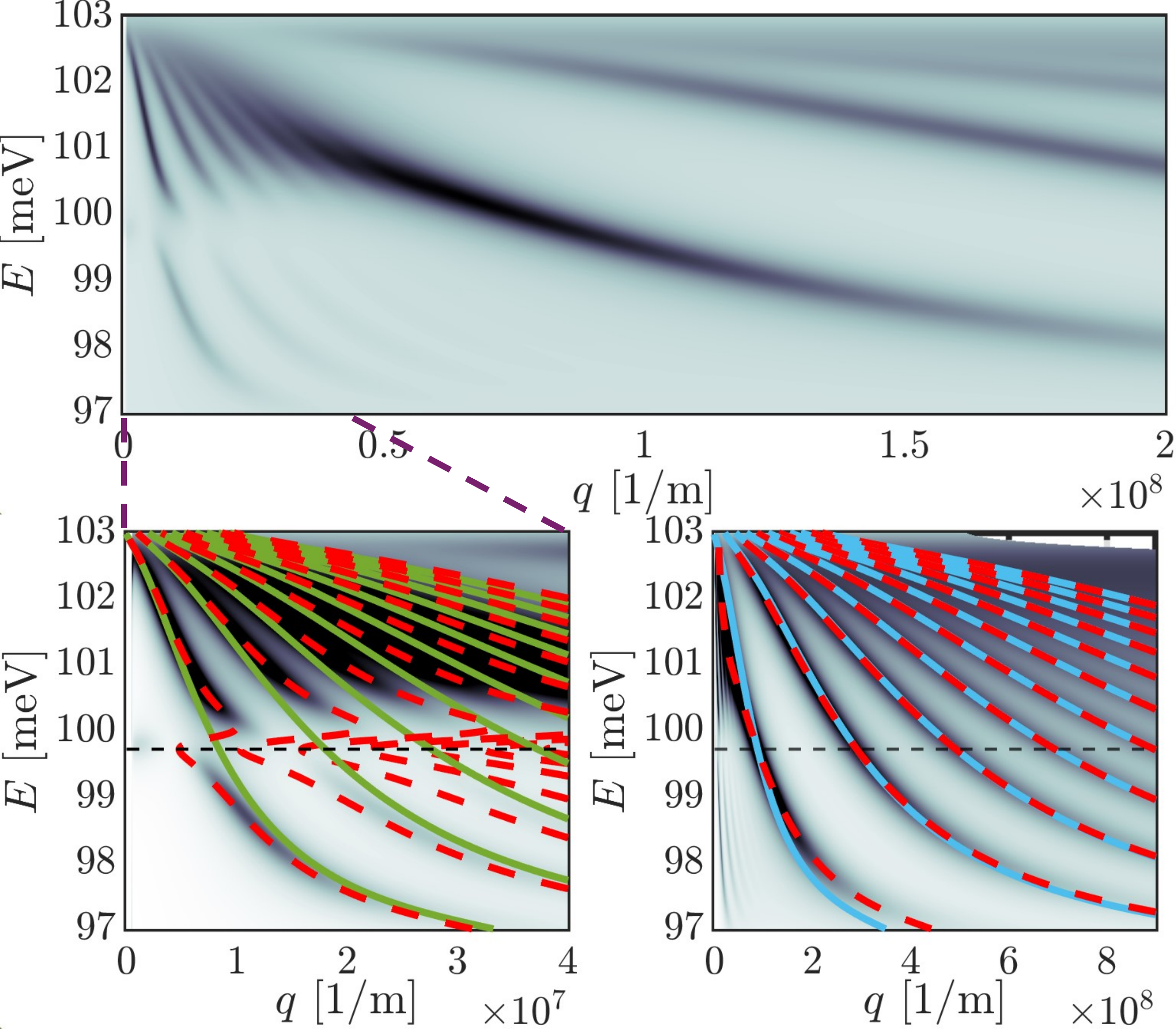}};
    \begin{scope}[x={(image.south east)},y={(image.north west)}]
    \node at (0,0.95) {(a)};
    \node at (0,0.45) {(b)};
    \node at (0.51,0.45) {(c)};
    \end{scope}
    \end{tikzpicture}
     \caption{Hybridization of BBLG excitons and hBN HPhP for the asymmetric structure in the lower Reststrahlen band, with $d_1=10\,\mathrm{nm}$ and $d_2=200\,\mathrm{nm}$. (a) The dispersion relation  of the hybridized polariton for BBLG with $V=58\,\mathrm{meV}$  simulated using TMM (colormap) showing two sets of modes. 
    (b) Zoom in on the low momentum modes of panel (a) together with the dispersion relation calculated from Eq. \ref{Eq hybrid asym low} (dashed red lines). Modes of HPhP for hBN with thickness of $d_1+d_2$ (green lines) and exciton energies of BBLG (dashed black lines) are also plotted.
    (c) Zoom out to the high momentum modes of panel (a) together with the dispersion relation calculated from Eq. \ref{Eq hybrid asym high} (dashed red lines). Odd modes of HPhP for hBN with thickness of $2d_1$ (blue lines) and exciton energies of BBLG (dashed black lines) are also plotted.
    The low momentum hybridized modes present an anticrossing behavior while the high momentum hybridized modes cross the main excitonic resonances.}
    \label{fig_hybrid_asym_lower}
\end{figure}
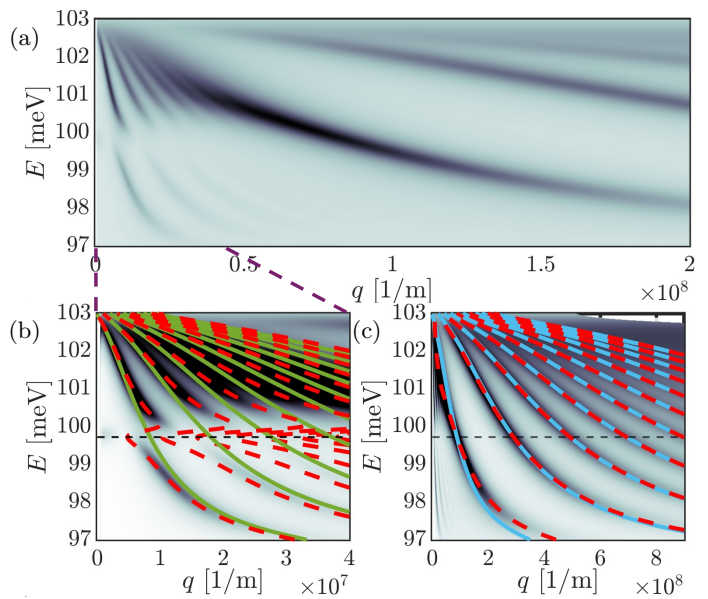

The second set of modes is the set of high momentum modes, which satisfies $|k_{hBN_z} d_{2,\sigma}| \gg 1$. The dispersion relation of these modes is given by:

\begin{align}    
    \begin{split}
    &    q_{Hybrid}^{high}=\\
    &=\frac{q_{HPhP}^{odd}(2d_1)}{2} +
    \sqrt{ \bigg(\frac{q_{HPhP}^{odd}(2d_1)}{2} \bigg)^2+\frac{i \omega \varepsilon_0 \varepsilon_{hBN_{zz}}}{\sigma d_1}}.
    \label{Eq hybrid asym high}
    \end{split}
\end{align}

The analytical solution for the high momentum modes given by Eq. \ref{Eq hybrid asym high} is presented in Fig. \ref{fig_hybrid_asym_upper}  (c) with dashed red lines.
Excellent agreement between the TMM simulation (colormap) and the analytical solution (dashed red line), obtained from Eq. \ref{Eq hybrid asym low} and \ref{Eq hybrid asym high}, can be seen in Fig. \ref{fig_hybrid_asym_upper}.

From Fig. \ref{fig_hybrid_asym_upper} (b), it is clear that the low momentum modes present an anti-crossing between the HPhP (green lines) and the exciton energies (dashed black lines). This anti-crossing is again an evidence of the hybridization.
The modes corresponding to $q^{high}$ (Eq. \ref{Eq hybrid asym high}), however, cross the main resonances of the BBLG (Fig. \ref{fig_hybrid_asym_upper} (c)), with their momentum being an order of magnitude higher than the $q^{low}$ (Eq. \ref{Eq hybrid asym low}) modes.

The dispersion relation in Eq. \ref{Eq hybrid asym high} does not depend on the thickness of the bottom hBN layer, since the conductivity of the BBLG plane is very large relative to the admittance of hBN, and the BBLG serves almost as a short circuit, therefore can be approximately described as a one-sided PEC (appendix \ref{Section app asym hyb}). 

The dispersion relations of the asymmetric structure in the lower Reststrahlen band are presented in Fig. \ref{fig_hybrid_asym_lower}, with $d_1=10\,\mathrm{nm}$ and $d_2=200\,\mathrm{nm}$. Similarly to the upper Reststrahlen band, the lower Reststrahlen band exhibits two sets of modes, distinguished by their lower and higher momentum.

\section{Conclusions}
 In conclusion, we have studies the hybridization properties of excitons and HPhPs in electrically tunable BBLG. We have derived the dispersion relations of these hybridized states, finding that the structure's symmetry plays a crucial role. Our results reveal the possible potential of the BBLG system as a tunable platform for observing strong light-matter interactions in the long wavelength regime. 

\section*{Acknowledments}
I.E. acknowledges the support of the European Union (ERC, TOP-BLG, Project No. 101078192). N.M.R.P acknowledges support from the European Union through project EIC PATHFINDER OPEN project No. 101129661-ADAPTATION, and  the Portuguese Foundation for Science and Technology (FCT) in the framework of the Strategic Funding UIDB/04650/2020, COMPETE 2020, PORTUGAL 2020, FEDER, and through project PTDC/FIS-MAC/2045/2021. Y.M. acknowledges funding from ISF grant 1089/22.

\newpage
\appendix

\section{Transmission Line Model}
\label{Section app TL}
We use the transmission line model to find the dispersion relations of the polaritonic modes. We can take the propagation direction to be $\hat{x}$ and assume uniformity of the fields in the $\hat{y}$ direction without loss of generality, due to symmetry in the $x-y$ plane in the structures investigated in our work. We can assume a transverse dependence of $e^{iqx}$ and assume a time dependence of $e^{-i\omega t}$, due to the continuity of the transverse electric field on the discontinuities planes of constant $z$ in the transmission between different materials, yielding the TM fields: $\mathbb{E}=(\hat{x}\mathbb{E}_x(z)+\hat{z}\mathbb{E}_z(z))e^{iqx}$ and $\mathbb{H}=\hat{y}\mathbb{H}_y(z)e^{iqx}$. The above transverse electric and magnetic fields, $\mathbb{E}_x(z)$ and $\mathbb{H}_y(z)$, satisfy the telegraph equations describing voltage and current in a transmission line \cite{DavidkChengField-and-wave-electromagnetics}, and will be referred to as voltage and current in the further discussion. The waves in the transmission line have a longitudinal dependence of $e^{\pm i k_{i_z} z}$, where $k_{i_z}$ is the $z$ component of the wavevector in the $i$th layer, related to the momentum $q$ by $\frac{q^2}{\varepsilon_{i_{zz}}}+\frac{k_{i_z}^2}{\varepsilon_{i_{xx}}}=k_0^2$, where $\varepsilon_{i_{xx}}$ and $\varepsilon_{i_{zz}}$ are the in-plane and out-of-plane permittivity of the $i$th layer and $k_0$ is the free-space wavenumber. These waves are referred to as forward and backward propagating waves.
The characteristic admittance of the $i$th layer is given by:
\begin{align}
    Y_i=\frac{\mathbb{H}^+_y(z)}{\mathbb{E}^+_x(z)}=-\frac{\mathbb{H}^-_y(z)}{\mathbb{E}^-_x(z)}=\frac{\omega \varepsilon_0 \varepsilon_{i_{xx}}}{k_{i_z}} ,
\end{align}
and describes the ratio between the current and the voltage of a forward propagating wave and equals to minus this ratio for a backward propagating wave in the $i$th transmission line.
The $z$ dependent admittance is defined as the ratio between the current and the voltage: $Y(z)=\frac{\mathbb{H}_y(z)}{\mathbb{E}_x(z)}$. If the $i$th layer is half infinite for $z \rightarrow \pm \infty$, then there is only a backward (or forward) propagating wave, and in this layer: $Y(z)=\pm Y_i$. Since polaritons are eigen modes of electromagnetic waves in the structure, their dispersion relation of can be found by solving the transverse resonance equation, obtained by equalizing the solutions of the $z$ dependent admittance at a point $z_0$ that satisfy the boundary conditions for $z>z_0$ and $z<z_0$. 
Assuming the graphene system in our structures is infinitesimal and located at $z=0$, the boundary condition at $z=0$ is given by: $\mathbb{H}_y(0^-)-\mathbb{H}_y(0^+)=\sigma \mathbb{E}_x(0)$, which in terms of admittances is given by: $Y(0^-)-Y(0^+)=\sigma$.

\section{Transverse resonance equation of HPhP}
\label{section app HPhP}
We now investigate the structure of air/hBN/air, where the thickness of the hBN is $d$. The $z$ dependent admittance at the air layers are given by $Y(z \geq \frac{d}{2})=Y_{air}$ and $Y(z \leq -\frac{d}{2})=-Y_{air}$, since the air layers are half infinite for $z \rightarrow \pm \infty$. 
The $z$ dependent admittance at the hBN layer is given by:
\begin{align}
    \label{Eq admittance using current}
    Y(z)=\frac{\mathbb{H}^+_y(z)+\mathbb{H}^-_y(z)}{\mathbb{E}^+_x(z)+\mathbb{E}^-_x(z)}=Y_{hBN}\frac{\mathbb{H}^+_y(z)+\mathbb{H}^-_y(z)}{\mathbb{H}^+_y(z)-\mathbb{H}^-_y(z)} .
\end{align}

Since the transverse field are continuous between layers, so is the $z$ dependent admittance. Therefore, the boundary condition at $z=\frac{d}{2}$ gives:

\begin{align}
    \label{Eq admittance BC}
    Y \bigg(\frac{d}{2} \bigg)=Y_{hBN}\frac{\mathbb{H}^+_y(\frac{d}{2})+\mathbb{H}^-_y(\frac{d}{2})}{\mathbb{H}^+_y(\frac{d}{2})-\mathbb{H}^-_y(\frac{d}{2})} = Y_{air}.
\end{align}

Eq. \ref{Eq admittance BC} gives the reflection coefficient for the current:

\begin{align}
    \label{Eq ref coef}
    r \bigg(\frac{d}{2} \bigg)=\frac{\mathbb{H}^-_y(\frac{d}{2})}{\mathbb{H}^+_y(\frac{d}{2})} = \frac{Y_{air}-Y_{hBN}}{Y_{air}+Y_{hBN}} .
\end{align}

Inserting the reflection coefficient and the $z$ dependence of the waves into \ref{Eq admittance using current} at $z=0$ gives the tangent equation:

\begin{align}
    \label{Eq admittance 0}
    \begin{split}
    Y(0) &=Y_{hBN} \frac{\mathbb{H}^+_y(\frac{d}{2}) e^{-i k_{hBN_z} \frac{d}{2}}+\mathbb{H}^-_y(\frac{d}{2}) e^{i k_{hBN_z} \frac{d}{2}}}{\mathbb{H}^+_y(\frac{d}{2}) e^{-i k_{hBN_z} \frac{d}{2}}-\mathbb{H}^-_y(\frac{d}{2}) e^{i k_{hBN_z} \frac{d}{2}}} = \\
    & = 
    Y_{hBN} \frac{Y_{air} - i Y_{hBN} tan(k_{hBN_z} \frac{d}{2})}{Y_{hBN} - i Y_{air} tan(k_{hBN_z} \frac{d}{2})} .
    \end{split}
\end{align}

Due to the symmetry of the structure around the $z$ axis, it supports even modes (even voltage around the $z$ axis) and odd modes (odd voltage around the $z$ axis). We note that the parity of the current is opposite to the parity of the voltage in TM polarization.
Since the transverse fields are continuous at $z=0$, the even modes satisfy a nullification of the current at $z=0$ and $Y(z=0)=0$ and the odd modes satisfy a nullification of the voltage at $z=0$ and $Y^{-1}(z=0)=0$.

The transverse resonance equation for the even modes is given by $Y_{air} - i Y_{hBN} tan(k_{hBN_z} \frac{d}{2})=0$ and for the odd modes is given by $Y_{hBN} - i Y_{air} tan(k_{hBN_z} \frac{d}{2})=0$. From the transverse resonance equations, the dispersion relation in Eq. \ref{Eq HPhP} can be obtained, with even modal orders for the even modes and odd modal orders for the odd modes.
 
\section{Transverse resonance equation of the Hybridized Modes}
\label{Section app hyb}
We now investigate a structure of air/hBN/BBLG/hBN/air. The $z$ dependent admittances at the air layers are given by $Y(z \geq d_2)=Y_{air}$ and $Y(z \leq -d_1)=-Y_{air}$, since the air layers are half infinite for $z \rightarrow \pm \infty$. 
Similarly to the derivation of Eq. \ref{Eq admittance 0}, the $z$ dependent admittance at $z=0^+$ in the bottom hBN layer is given by:

\begin{align}
    \label{Eq admittance 0+}
    Y(0^+) =
    Y_{hBN} \frac{Y_{air} - i Y_{hBN} tan(k_{hBN_z} d_2)}{Y_{hBN} - i Y_{air} tan(k_{hBN_z} d_2)} .
\end{align}

The values of $d_1$ and $d_2$ do not have strict limitations generally, although we will derive analytical expression for two different cases, which limit the value of thicknesses for our discussion. 

Similarly, the $z$ dependent admittance at $z=0^-$ is given by:

\begin{align}
    \label{Eq admittance 0-}
    Y(0^-) &=
    Y_{hBN} \frac{Y_{air} - i Y_{hBN} tan(k_{hBN_z} d_1)}{Y_{hBN} - i Y_{air} tan(k_{hBN_z} d_1)} .
\end{align}

By inserting the relations in Eq. \ref{Eq admittance 0+} and \ref{Eq admittance 0-} into the boundary condition at $z=0$, we obtain the transverse resonance equation:

\begin{align}  
    \begin{split}
    &
    Y_{hBN} \frac{Y_{air}-iY_{hBN}tan(k_{hBN_{z}}d_1)}{Y_{hBN}-iY_{air}tan(k_{hBN_{z}}d_1)} +\\
    & +
    Y_{hBN} \frac{Y_{air}-iY_{hBN}tan(k_{hBN_{z}}d_2)}{Y_{hBN}-iY_{air}tan(k_{hBN_{z}}d_2)} +
     \sigma =0 .  
     \end{split}
    \label{Eq hyb admittance matching}      
\end{align}  

\section{Dispersion Relation of the Symmetric Hybridized Modes}
\label{Section app sym hyb}

The odd modes of the hybridized polaritons in the symmetric case satisfy a nullification of the voltage at $z=0$ and $Y^{-1}(z=0)=0$. Therefore, they are obtained by equalizing the denominator Eq. \ref{Eq admittance 0+} or \ref{Eq admittance 0-} to $0$, which gives the transverse resonance equation of odd modes of HPhPs.

The even modes of the hybridized polaritons in the symmetric case do not satisfy a nullification of the current at $z=0$ due to the presence of the graphene system. Therefore, we cannot use the simplification of the symmetric structures here.
The transverse resonance equation of even modes is obtained by setting $d_1=d_2=\frac{d}{2}$ in Eq. \ref{Eq hyb admittance matching}. To derive the dispersion relation we use the trigonometric identity $tan(\theta_1+\theta_2)=\frac{tan(\theta_1)+tan(\theta_2)}{1-tan(\theta_1)tan(\theta_2)}$ with $\theta_1=k_{hBN_{z}}\frac{d}{2}$ and $\theta_2=\arctan (i \frac{Y_{air}}{Y_{hBN}})$ on the transverse resonance equation and obtain:

\begin{align}  
    tan \bigg(k_{hBN_{z}}\frac{d}{2}+\arctan \bigg(i \frac{Y_{air}}{Y_{hBN}} \bigg) \bigg) =
     \frac{\sigma}{2iY_{hBN}} . 
    \label{Eq hyb admittance matching even}      
\end{align}  

We now use the $\arctan$ operator on both sides of Eq. \ref{Eq hyb admittance matching even} and use the approximation $\arctan(\frac{\sigma}{2iY_{hBN}}) \approx \frac{\sigma}{2iY_{hBN}}$ due to assumption of $|\frac{\sigma}{Y_{hBN}}| \ll 1$:

\begin{align}  
    k_{hBN_{z}}\frac{d}{2}+\arctan \bigg(i \frac{Y_{air}}{Y_{hBN}} \bigg) + \pi L_{even} =
     \frac{\sigma}{2iY_{hBN}} , 
    \label{Eq hyb admittance matching even 2}      
\end{align} 

where $L_{even}$ is an integer that satisfies $|\frac{\sigma}{Y_{hBN}}| \ll 1$.
We now define $d_\sigma=\frac{i\sigma}{\omega \varepsilon_0 \varepsilon_{hBN_{xx}}}$ and write $\frac{\sigma}{Y_{hBN}}=-ik_{hBN_{z}}d_\sigma$.
 Using the approximation $q \gg \sqrt{\varepsilon_{hBN_{zz}}}k_0$, we can write $k_{hBN_z}\approx i q \sqrt{\frac{\varepsilon_{hBN_{xx}}}{\varepsilon_{hBN_{zz}}}}$ and $k_{air_z}\approx i q$, and therefore $\frac{Y_{air}}{Y_{hBN}}=\varepsilon_{hBN_{eff}}^{-1}$. Setting these relations in Eq. \ref{Eq hyb admittance matching even 2} gives the dispersion relation of the even hybridized modes in Eq. \ref{Eq hybrid sym even}.
 For the parameters of Fig. \ref{fig_hybrid_sym} (a), the lower Reststrahlen band only supports one even mode, with modal order $L_{even}=-1$ which corresponds to $L=2L_{even}=-2$ of HPhP. This is the only even mode that solves Eq. \ref{Eq hyb admittance matching} with the parameters of this case since the assumption of $|\frac{\sigma}{Y_{hBN}}| \ll 1$ does not hold for the other modes.

 The analytical dispersion relation in Eq. \ref{Eq hybrid sym even} was derived under the approximation of $|k_{hBN_{z}}d_\sigma| \ll 1$, which can be written as $|\frac{\sigma}{Y_{hBN}}| \ll 1$: the BBLG plane has a very low admittance relative to the hBN, and therefore the influence of the BBLG can be considered as a small correction to the dispersion relation of the HPhPs.

Far from the excitonic resonances of the BBLG, $d_\sigma$ is negligible, and the even hybridized modes converge to even modes of HPhPs. Near the excitonic resonances, the contribution of the conductivity is significant, and $d_\sigma$ effectively increases or decreases the thickness of the hBN layers, according to the signs of the conductivity of BBLG and of the in-plane permittivity of hBN. In the lower Reststrahlen band $Re(\varepsilon_{hBN_{xx}})>0$ (Fig. \ref{fig_cond} (a)), therefore below the main excitonic resonance of BBLG where $Im(\sigma)<0$: $d_\sigma>0$ (Fig. \ref{fig_hybrid_dsigma} (a)) and $d_{eff}>d$ and above the main excitonic resonance of BBLG where $Im(\sigma)>0$: $d_\sigma<0$ and $d_{eff}<d$. In the upper Reststrahlen band $Re(\varepsilon_{hBN_{xx}})<0$ (Fig. \ref{fig_cond} (a)), therefore below the main excitonic resonance of BBLG $d_{eff}<d$ and above it $d_{eff}>d$. From Eq. \ref{Eq HPhP} it is clear that the momentum of HPhP is inverse to the thickness of the hBN layer. Hence, when $d_{eff}<d$, the even hybridized modes have an increased momentum compared to the HPhPs, and when $d_{eff}>d$, the even hybridized modes have a decreased momentum compared to the HPhPs (Fig. \ref{fig_hybrid_sym}). The discussion above is summarized in table \ref{table anticross}.

\begin{table}[htbp]
  \centering
  \begin{tabular}{|c|c|c|c|c|c|}
    \hline
     &  & $Re(\varepsilon_{hBN_{xx}})$ & $Im(\sigma)$ & \makecell{$Re(d_{\sigma})=$ \\ $Re(d_{eff}) - d$} & \makecell{$Re(q_{Hybrid}^{even}$ \\ $-q_{HPHp}^{even})$} \\
    \hline
    \multirow{2}{*}{\rotatebox{90}{\makecell{Lower \\ Reststrahlen \\ band}}} & \makecell{Below \\ main \\ res.} & $+$ & $-$ & $+$ & $-$\\
    \cline{2-6}
    & \makecell{Above \\ main \\ res.} & $+$ & $+$ & $-$ & $+$\\
    \hline
    \multirow{2}{*}{\rotatebox{90}{\makecell{Upper \\ Reststrahlen \\ band}}} & \makecell{Below \\ main \\ res.} & $-$ & $-$ & $-$ & $+$\\
    \cline{2-6}
    & \makecell{Above \\ main \\ res.} & $-$ & $+$ & $+$ & $-$\\
    \hline
  \end{tabular}
  \caption{Explanation of the anticrossing behavior of the even modes of the hybridized polariton in the symmetric structure. The $+$ and $-$ signs describe the signs of the quantities in the different spectral regions.}
  \label{table anticross}
\end{table}

 In the upper reststrahlen band, the even hybridized modes do not converge with the even HPhPs at high energies, as can be seen for the $L=0$ mode in Fig. \ref{fig_hybrid_sym} (a). Although these energies are far from the excitonic resonances, they are near the in-plane longitudinal optical phonon frequency of the hBN, where the in-plane permettivity of hBN is very small and therefore $d_{\sigma}$ is not negligible.

\section{The additional polarization current}
\label{Section app d_sigma}
In order to understand the meaning of $d_\sigma$, we use Ampere's law in the $x$ direction around $z=0$:

\begin{equation}
    -\partial_z \mathbb{H}_{y} = - i \omega \varepsilon_0 \varepsilon_{hBN_{xx}} \mathbb{E}_x + J ,
    \label{Eq Amperes law}
\end{equation}

where $J$ is the current density due to the presence of the BBLG. The surface current density on the BBLG  $J_s = J \cdot d_{BBLG}$ is given by Ohm's law $J_s = \sigma \mathbb{E}_x$, 
where $d_{BBLG}$ is the 
thickness of the BBLG. Therefore, Eq. \ref{Eq Amperes law} can be written as: 

\begin{equation}
    -\partial_z \mathbb{H}_{y} =- i \omega \varepsilon_0 \varepsilon_{hBN_{xx}} \bigg(1+\frac{d_\sigma}{d_{BBLG}}  \bigg)\mathbb{E}_x .
    \label{Eq Amperes law with dsigma}
\end{equation}

The parameter $d_\sigma$ therefore provides the additional polarization current that flows on the BBLG in terms of length.

\begin{figure}[t]
    \centering
    \hspace*{-13pt} 
    \begin{tikzpicture}
    \node[anchor=south west,inner sep=0] (image) 
    at (0,0) 
    {\includegraphics[width=1.03\columnwidth]{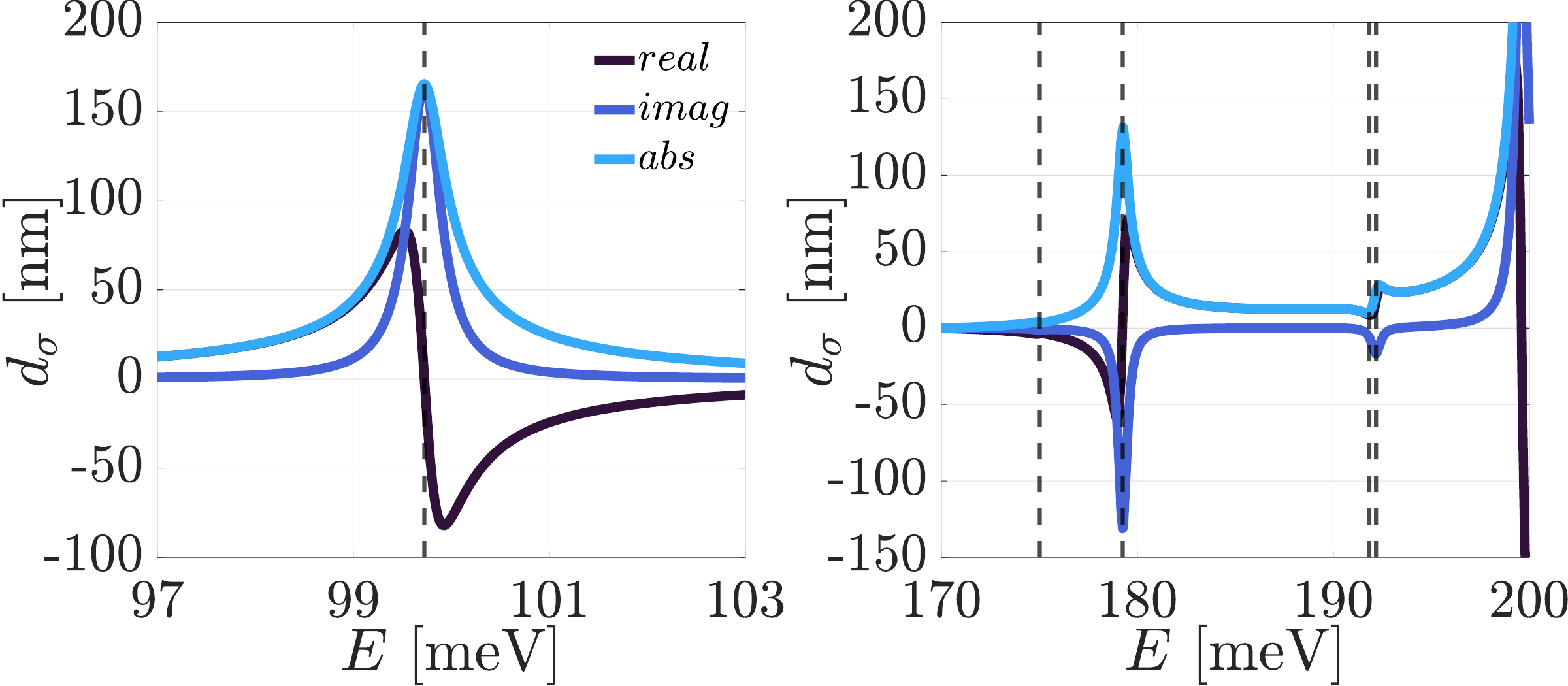}};
    \begin{scope}[x={(image.south east)},y={(image.north west)}]
    \node at (0,0.95) {(a)};
    \node at (0.5,0.95) {(b)};
    \end{scope}
    \end{tikzpicture}
    \caption{
    $d_\sigma$ for (a) $V=58\,\mathrm{meV}$ and (b) $V=115\,\mathrm{meV}$.}
    \label{fig_hybrid_dsigma}
\end{figure}

The frequency dependence of $d_\sigma$ is presented in Fig \ref{fig_hybrid_dsigma} (a) and (b), for the lower Reststrahlen band with $V=58\,\mathrm{meV}$ and for the upper Reststrahlen band with $V=115me\,\mathrm{meV}$, respectively.

\section{Dispersion Relation of the Asymmetric Hybridized Modes}
\label{Section app asym hyb}

Under the chosen conditions of $d_1, |d_\sigma| \ll d_2$, the transverse resonance equation has two sets of solutions.
The derivation of the set of low momentum modes is similar to the derivation of the even modes in the symmetric case. 
Similarly to Eq. \ref{Eq hyb admittance matching even} we can write:

\begin{align}  
    \begin{split}
    &tan \bigg(k_{hBN_{z}}d_2+\arctan \bigg(i \frac{Y_{air}}{Y_{hBN}} \bigg) \bigg) \\=
    &-tan \bigg(k_{hBN_{z}}d_1+\arctan \bigg(i \frac{Y_{air}}{Y_{hBN}} \bigg) \bigg)+
     \frac{\sigma}{iY_{hBN}} . 
    \end{split}
    \label{Eq hyb admittance matching low}      
\end{align}

We now use the $\arctan$ operator on both sides of Eq. \ref{Eq hyb admittance matching low} and use the approximation $\arctan(x) \approx x$ with $x$ being the right-hand-side of Eq. \ref{Eq hyb admittance matching low}, due to assumptions of $|\frac{\sigma}{Y_{hBN}}| \ll 1$ and $|k_{hBN_{z}}d_1| \ll 1$. 
We then use the same approximation with $x=k_{hBN_{z}}d_1+\arctan (i \frac{Y_{air}}{Y_{hBN}})$, and obtain:

\begin{align}  
    k_{hBN_{z}}(d_1+d_2)+2\arctan \bigg(i \frac{Y_{air}}{Y_{hBN}} \bigg) + \pi L_{low} =
     \frac{\sigma}{iY_{hBN}} . 
    \label{Eq hyb admittance matching low 2}      
\end{align} 

Using $\frac{\sigma}{Y_{hBN}}=-ik_{hBN_{z}}d_\sigma$, $\frac{Y_{air}}{Y_{hBN}}=\varepsilon_{hBN_{eff}}^{-1}$ and $k_{hBN_z}\approx i q \sqrt{\frac{\varepsilon_{hBN_{xx}}}{\varepsilon_{hBN_{zz}}}}$ we obtain the dispersion relation of the low modes in Eq. \ref{Eq hybrid asym low}.

To derive the dispersion relation of the set of high momentum modes we start by using the assumption that $d_2$ is larger than all quantities with length dimensions, therefore the bottom layer can be approximated as half infinite and $Y(0^+)=Y_{hBN}$. Using the assumption of $|\frac{\sigma}{Y_{hBN}}| \gg 1$, we can neglect $Y(0^+)$, and the transverse resonance equation is therefore $Y(0^-)+\sigma=0$, where $Y(0^-)$ is given by Eq. \ref{Eq admittance 0-}. 
Now, similarly to Eq. \ref{Eq hyb admittance matching even} we can write:

\begin{align}  
    tan \bigg(k_{hBN_{z}}d_1+\arctan \bigg(i \frac{Y_{air}}{Y_{hBN}} \bigg) \bigg) =
     \frac{\sigma}{iY_{hBN}} . 
    \label{Eq hyb admittance matching high}      
\end{align}

We now use the $\arctan$ operator on both sides of Eq. \ref{Eq hyb admittance matching high} and use the approximation $\arctan(\frac{\sigma}{iY_{hBN}}) \approx \frac{\pi}{2}-\frac{iY_{hBN}}{\sigma}$ due to assumption of $|\frac{\sigma}{Y_{hBN}}| \gg 1$:

\begin{align}  
    k_{hBN_{z}}d_1+\arctan \bigg(i \frac{Y_{air}}{Y_{hBN}} \bigg) + \pi L_{high} =
     \frac{\pi}{2}-\frac{iY_{hBN}}{\sigma} . 
    \label{Eq hyb admittance matching high 2}      
\end{align} 

Using $\frac{\sigma}{Y_{hBN}}=-ik_{hBN_{z}}d_\sigma$ and $\frac{Y_{air}}{Y_{hBN}}=\varepsilon_{hBN_{eff}}^{-1}$, and multiplying Eq. \ref{Eq hyb admittance matching high 2} by $k_{hBN_{z}}$, we obtain a quadratic equation in terms of $k_{hBN_{z}}$:

\begin{align}  
    k_{hBN_{z}}^2 d_1-k_{hBN_{z}} \cdot k_{hBN_{z}}^{HPhP,odd} (2d_1) \cdot d_1 +
     (d_\sigma)^{-1} =0 , 
    \label{Eq hyb admittance matching high 3}      
\end{align} 

with

\begin{align}    
    \begin{split}
    &k_{hBN_{z}}^{HPhP,odd} (2d_1) =\\
    &=-\frac{1}{2d_1} \bigg( 2\arctan \bigg(\frac{i}{\varepsilon_{hBN_{eff}}} \bigg)
    + \pi (2L_{high}-1) \bigg) ,         
    \end{split}
    \label{Eq odd 2d1}
\end{align}

where $k_{hBN_{z}}^{HPhP,odd} (2d_1)$ is the $z$ component of the wavevector in hBN for odd modes of HPhP with hBN with thickness $2d_1$. Solving the above equation and setting $k_{hBN_z}\approx i q \sqrt{\frac{\varepsilon_{hBN_{xx}}}{\varepsilon_{hBN_{zz}}}}$ gives the dispersion relation of the high modes in Eq. \ref{Eq hybrid asym high}.

The dispersion relation in Eq. \ref{Eq hybrid asym high} does not depend on the thickness of the bottom hBN layer. Since $d_2$ is much larger than all quantities with length dimensions, we can assume that the thicker hBN slab is half infinite, making its contribution negligible. Another explanation is that the conductivity of the BBLG plane is very large relative to the admittance of hBN since the condition $|k_{hBN_z} d_{\sigma}| \gg 1$ can be written as $|\frac{\sigma}{Y_{hBN}}| \gg 1$, and the BBLG serves almost as a short circuit, therefore can be approximately described as a one-sided PEC. 
For high values of the conductivity and for high modal orders, the assumption $|k_{hBN_z} d_{\sigma}| \gg 1$ is more accurate, and the hybridized modes converge to odd modes of HPhP with thickness $2d_1$: $q_{Hybrid}^{high}=q_{HPhP}^{odd} (2d_1)$, and are not affected by the BBLG (Fig. \ref{fig_hybrid_asym_lower} (c) and Fig. \ref{fig_hybrid_asym_upper} (c)). This can be understood through the validity of the approximation of the BBLG as a PEC, which divides the structure into two areas. For $z>0$ there is half infinite hBN which cannot support guided modes. For $z<0$ there is a hBN slab with thickness $d_1$ with air on one side and a high conductivity on the other. In this area, the solution is equivalent to a finite slab with $2d_1$ where the transverse electric field is zero at $z=0$ - odd modes.
The high momentum modes present around the main resonance an opposite behavior compared to the anticrossing observed for the low momentum modes. These modes exist in the crossing between the HPhP and the main excitonic resonances, where the conductivity is high and the hybridized modes converge to HPhPs. At energies far from the main resonance, where the conductivity is lower, the influence of the BBLG is more significant.

 Nevertheless, it can be seen in Fig. \ref{fig_hybrid_asym_upper} (c) that around the additional resonance at $\approx 192\,\mathrm{meV}$ an "anticrossing" behavior occurs (similar to \ref{fig_hybrid_sym} (a) for $L=2$ at the same energy), compared to the crossing of the main resonances. Again, this is not an exact anticrossing behavior since the hybridized modes remain on the same side of the HPhP dispersion curves. To understand this phenomenon, we should examine Eq. \ref{Eq hybrid asym high} around the resonances. Near the main resonance, the imaginary part of the BBLG conductivity changes signs (Fig. \ref{fig_cond} (b)), so the hybridized modes added momentum, with respect to the HPhP, also changes its sign when the energy is below or above the main resonance, resulting in a crossing of the HPhP dispersion. Near the higher-order resonances, the imaginary part of the conductivity remains positive, yielding a larger momentum for the hybridized mode compared with HPhP both above and below the resonance. Around this resonance there is, however, a sharp transition in the value of $\sigma$, resulting in the "anticrossing" behavior.


\bibliography{ExPhHy}  

\end{document}